\documentclass{aa}

\usepackage{txfonts}

\usepackage{graphicx}
\usepackage{ifthen}
\usepackage{url}
\usepackage{amsmath}
\usepackage{bm}
\usepackage{lscape}
\usepackage{rotating}
\usepackage[graphicx]{realboxes}
\usepackage{supertabular}
\usepackage{pdfpages}
\usepackage{footnote}
\usepackage{hyperref} 
\usepackage{amssymb}
\usepackage{multirow}
\usepackage{array}
\usepackage{xcolor}

\definecolor{myurlcolor}{HTML}{3B54A3}

\hypersetup{
  colorlinks   = true,    
  urlcolor     = myurlcolor,    
  linkcolor    = myurlcolor,    
  citecolor    = myurlcolor,      
  filecolor   = myurlcolor
}

\DeclareMathOperator*{\argmin}{argmin}

\begin{document}
\title{\texttt{EROSE}: An algorithm searching for resolved Local Group satellites}
\subtitle{The impact of combining different photometric bands on the detectability of Milky Way satellites in the LSST}

\author{Samuel Rusterucci \inst{1, 2}, Nicolas F. Martin \inst{1, 3}, Else Starkenburg \inst{2}}

\institute{Université de Strasbourg, CNRS, Observatoire astronomique de Strasbourg, UMR 7550, F-67000 Strasbourg, France; \\ \email{samuel.rusterucci@astro.unistra.fr} \and Kapteyn Astronomical Institute, University of Groningen, Landleven 12, 9747 AD Groningen, The Netherlands \and Max-Planck-Institut für Astronomie, Königstuhl 17, D-69117 Heidelberg, Germany}
\date{Received 25 March 2026 / Accepted 8 August 2026}

\titlerunning{\textsc{erose}: a Local Group satellite search algorithm}
\authorrunning{Rusterucci et al.}

\abstract{
Systematic searches for ultra-faint Milky Way satellite galaxies essentially consist of local density estimators coupled with colour-magnitude cuts following stellar isochrone tracks. These selections are typically performed in 2D colour-magnitude spaces and have rarely, if ever, been extended to higher-dimensional spaces. Moreover, these studies typically only consider $g$-, $r$-, and $i$-band photometry, disregarding the use of the shallower but metal-sensitive $u$ band. Entering a new era of wide-field surveys, we focused our study on LSST, which will provide exceptional depth and uniform coverage in all $ugri$ bands. We conducted our analysis using simulated data from the LSST Data Challenge 2 (DC2), which covers a $300~\mathrm{deg}^2$ high-galactic-latitude region of the sky and reaches the expected survey depth after five years of observations. After generating and injecting simulated dwarf galaxies of various properties into the DC2 footprint, we attempted to retrieve them using \textsc{erose}, a new fast multi-band searching algorithm designed for easy use in any survey and any combination of input photometric bands. Exploring different combinations of photometric bands, we find that a multi-colour magnitude space combining all three $gri$ photometric bands slightly outperforms, on average, the more commonly used colour-magnitude spaces combining only $gr$ or $ri$. However, the highest recovery fraction is achieved using a colour-magnitude space combining the $u$ and $g$ bands, recovering on average $64.1^{+6.3}_{-6.0}\%$ of the injected satellites, a value slightly higher than that obtained when combining $gri$. This improvement is particularly evident for the most extended systems, for which the additional colour information provided by the $u$ band improves our ability to isolate metal-poor member stars from contaminating foreground populations. We estimate that the LSST will discover $\sim 50$ new Milky Way dwarf galaxies, more than doubling the currently known number of satellites within its footprint. Combining the LSST photometric bands with a space-based star--galaxy separation could further improve these results, enabling discoveries of satellites populating the low-mass end of the Milky Way's satellite galaxy luminosity function and providing valuable constrains on dark matter models.
}

\maketitle
\nolinenumbers

\section{Introduction}
Studies by \citet{Aaronson_1983} and \citet{Faber_1983} were the first to show that dwarf galaxies contain a substantial dark matter component, a property so fundamental to defining those objects that it became an essential part of their recognised definition \citep[see][]{Willman_2012}. Today, dwarf galaxies are known to be amongst the most dark-matter-dominated objects in the Universe \citep{Simon_2011} and despite their -- at times -- ambiguous nature, the mere existence of these faint systems orbiting the Milky Way can place constraints on the nature of dark matter \citep[][]{Lovell_2012, Nadler_2021} or provide upper bounds on the lowest-mass halos \citep{Jethwa_2017}. These reasons justify past and ongoing endeavours to detect and study ever-fainter satellites \citep{Doliva_Dolinsky_2026}.

The turning point in these searches came with the advent of wide-field digital photometric surveys at the start of the 21$^{\mathrm{st}}$ century. Prior to this, only a handful of Milky Way companions were known, almost all discovered visually through meticulous analyses of photographic plates. Automated, more efficient (and less tiring) methods were only truly introduced along with the Sloan Digital Sky Survey (SDSS) and led to the discovery of a slew of Milky Way companions \citep[e.g.][]{Willman_2005, Zucker_2006, Belokurov_2007}. When applied to the Panoramic Survey Telescope and Rapid Response System 3$\pi$ survey (Pan-STARRS1, PS1), the Dark Energy Survey (DES), and the DECam Local Volume Exploration Survey (DELVE), these methods led to even more discoveries \citep[e.g.][]{Laevens_2015, Koposov_2015, Drlica_Wagner_2015, Cerny_2023}, increasing the total number of known Milky Way satellites sixfold compared with the previous millennium.

Despite the increased depth of the data, the underlying principles of these search algorithms \citep[e.g.][]{Koposov_2008, Bechtol_2015} have remained unchanged, continuing to rely on kernel-based density estimation combined with 2D colour–magnitude selections following empirical or theoretical isochrone tracks. These photometric selections are implemented using a combination of 1) deep bands and 2) colours that best distinguish stars in (metal-poor) dwarf galaxies from (metal-rich) Milky Way foreground contamination. Therefore, it seems natural to include as many photometric bands as possible (not only two) and, in particular, the metal-sensitive $u$ band. This band was used in the past to produce metallicity maps of the Milky Way \citep[e.g.][]{Ivezic_2008, Ibata_2017}. However, to our knowledge, it was never systematically included in dwarf-galaxy searches, as it is typically shallower than the more commonly used $g$, $r$, and $i$ bands.

Entering the era of the Legacy Survey of Space and Time (LSST), the purpose of our study is to highlight the untapped potential of the $u$ band in this search for new Milky Way dwarf galaxy satellites. This paper is structured as follows. In Sect.~\ref{sec:search_algorithm}, we introduce \textsc{erose}, our new automated and highly modular search algorithm, able to process the entire LSST footprint in a matter of hours on a laptop using any combination of input photometric bands. In Sect.~\ref{sec:data} we present the LSST DC2 simulated sky used in our study along with our methodology for generating and injecting dwarf galaxies. Our methodology includes a catalogue-level approximation for source blending, reproducing trends obtained by \citet{Zhang_2025}, who injected dwarf galaxies at the image level. Sect.~\ref{sec:results} presents our recovery fraction as a function of distance, absolute magnitude and size of our input dwarf galaxies. Using the empirical model of the Milky Way satellite population from \citet{Tan_2025}, we estimated the number of satellite discoveries expected within the LSST footprint. Our results are consistent with those of \citet{Tsiane_2025}, and we discuss this comparison in Sect.~\ref{sec:discussion}, along with our conclusions.


\section{\textsc{erose}: A new search algorithm}
\label{sec:search_algorithm}
\subsection{Overview}
We present \textsc{erose}\footnote{The algorithm is entirely Python-based and publicly available at \url{https://github.com/samuelrusterucci/erose}} (Enhancing Rapidly Overdensities of Sources with Ease), a new automated search algorithm based on the principles of a methodology already proven efficient for detecting faint Local Group satellites in wide-field surveys \citep[e.g.][]{Koposov_2015, Bechtol_2015, Smith_2023}. This algorithm expands upon previous work by incorporating a (multi-)colour-magnitude statistical weighting of stars, along with a background estimate returning a minimal number of contaminants. The algorithm is fast, entirely Python-based and requires minimal inputs, making it easily adaptable to any survey with only minor adjustments. 

\begin{figure*}
	\centering
	\includegraphics[width=\hsize]{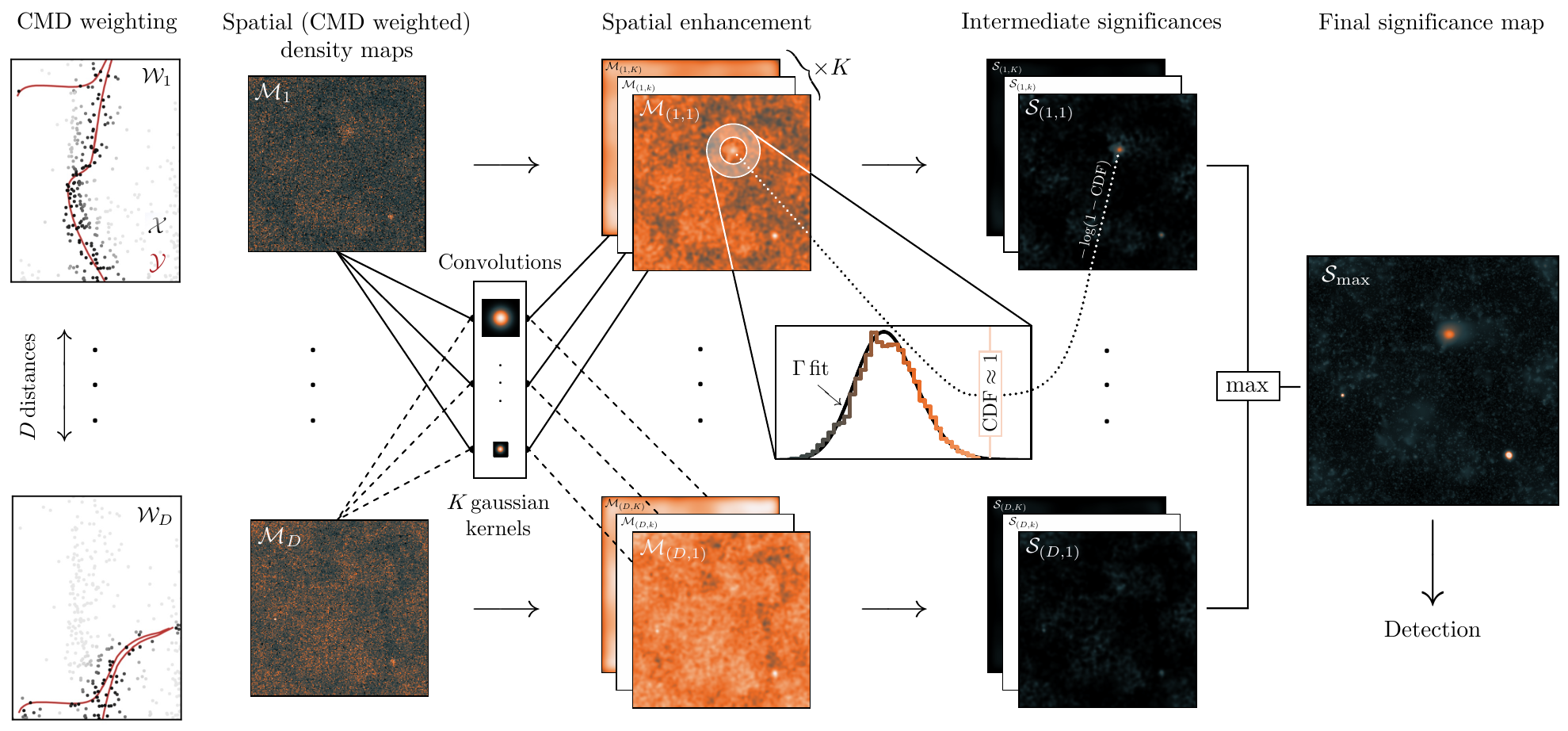}
	\caption{Schematic representation of the \textsc{erose} algorithm.}
	\label{fig:how_erose}
\end{figure*}

Before presenting the more mathematical description of \textsc{erose} in Sects.~\ref{sub_sec:isochrone_selection} to \ref{sub_sec:significance_maps}, it is worth understanding its basic principles, summarised in Fig.~\ref{fig:how_erose}. The algorithm begins by weighting stars based on their consistency with a theoretical or empirical isochrone, a procedure repeated for different distance shifts. After spatially binning the weighted stars, we obtain a density map for each distance. \textsc{erose} convolves these maps with Gaussian kernels of different sizes with the aim of enhancing overdensities of varying angular scales. To evaluate the significance of a given pixel in any of these maps, a gamma distribution is fitted to the distribution of neighbouring pixels (contained within a surrounding annulus), and the value of the cumulative distribution function (CDF) of the central pixel is used as part of our significance metric. Repeating this procedure for every pixel of every map, we obtain our intermediate significance maps. Finally, by taking the maximum pixel value across all maps, we obtain our final significance map, from which overdensities can be extracted and characterised.

\subsection{Weighting stars in the (multi-)colour-magnitude diagram}
\label{sub_sec:isochrone_selection}
For a given combination of $C$ photometric bands $\{b_z\}_{z=1}^C$, one can define a $C$-dimensional (multi-)colour-magnitude space, where each dimension is denoted as $\zeta_c$. Here, $\{\zeta_c\}_{c=1}^{C - 1}$ represent colours (i.e. linear combinations of magnitudes in two bands), and $\zeta_C$ is a magnitude\footnote{For example, we could have ($b_1, b_2$) = ($g, r$) and ($\zeta_1, \zeta_2$) = ($g-r, g$), where $g$ and $r$ are the LSST photometric bands.}. For a stellar field of arbitrary size and centre, \textsc{erose} assigns each star a weight based on its likelihood of being consistent with a stellar population modelled by a theoretical isochrone in the same $C$-dimensional space. Let $\mathcal{X} = \{ \mathbf{x}_i\}_{i=1}^I$ be the set of $I$ data points representing the colours and magnitudes of the $I$ input stars. Likewise, let $\mathcal{Y} = \{ \mathbf{y}_j \}_{j =1}^J$ be the set of $J$ points modelling the isochrone.

Assuming the per-band uncertainties are known as a function of magnitude (see Sect.~\ref{sub_sec:uncert_completeness}), the algorithm assigns a $C\times C$ variance-covariance matrix to each $\mathbf{y}_j$:
\begin{eqnarray}
	\tilde{\boldsymbol{\Sigma}}_j = 
	\begin{pmatrix}
		\sigma_{b_1, j}^2 + p_{b_1}^2 & \cdots & 0\\
		\vdots & \ddots & \vdots\\
		0 & \cdots & \sigma_{b_C, j}^2  + p_{b_C}^2
	\end{pmatrix}.
	\label{eq:per_bands_phot_uncert}
\end{eqnarray}
Each of its diagonal elements is the sum of squares of the photometric uncertainties at $\mathbf{y}_j$ in the $b_z$ band, denoted $\sigma_{b_z, j}$, and an additional width term $p_{b_z}$. We note that the off-diagonal elements are null because we assume the measurements are independent in the different bands\footnote{This is a reasonable assumption for most surveys, although it is not strictly valid for LSST (details in Sect.~\ref{sub_sec:uncert_completeness}).}. In contrast, our $C$-dimensional (multi-)colour-magnitude space inherently exhibits correlations, since individual bands may be shared among different $\zeta_c$ axes. We propagate the individual per-band uncertainties encoded in $\tilde{\boldsymbol{\Sigma}}_j$ to each dimension of this space through
\begin{eqnarray}
	\boldsymbol{\Sigma}_j = \boldsymbol{J} \tilde{\boldsymbol{\Sigma}}_j \boldsymbol{J}^T,
	\label{eq:covariance_mcms} 
\end{eqnarray}
where $\boldsymbol{J}$ is the Jacobian matrix defined as
\begin{eqnarray}
	\boldsymbol{J} = 
	\begin{pmatrix}
		\displaystyle \frac{\partial \zeta_1}{\partial b_1} & \cdots & \displaystyle \frac{\partial \zeta_1}{\partial b_\mathcal{C}}\\
		\vdots & \ddots & \vdots\\
		\displaystyle \frac{\partial \zeta_\mathcal{C}}{\partial b_1} & \cdots & \displaystyle \frac{\partial \zeta_\mathcal{C}}{\partial b_\mathcal{C}}
	\end{pmatrix}.
	\label{eq:jacobian}
\end{eqnarray}

Finally, we can define the mapping $f:\mathcal{X} \rightarrow \mathcal{Y}$ such that each $\mathbf{x}_i$ is assigned to its nearest neighbour in $\mathcal{Y}$:
\begin{eqnarray}
	f(\mathbf{x}_i) = \underset{\mathbf{y}_j \in \mathcal{Y}}{\argmin} || \mathbf{x}_i - \mathbf{y}_j ||,
	\label{eq:mapping}
\end{eqnarray}  
where in other words, $f(\mathbf{x}_i)$ provides the index of the point of $\mathcal{Y}$ that minimises the distance to the point $\mathbf{x}_i$. Following this formalism, the weight associated to each star is
\begin{eqnarray}
	w_i \propto \exp \Bigg(-\frac{1}{2} \big(\mathbf{x}_i - \mathbf{y}_{f(\mathbf{x}_i)}\big)^\mathrm{T} \boldsymbol{\Sigma}^{-1}_{f(\mathbf{x}_i)} \big(\mathbf{x}_i - \mathbf{y}_{f(\mathbf{x}_i)}\big)\Bigg),
	\label{eq:distance_isochrone}
\end{eqnarray}  
and measures its likelihood to be consistent with a given isochrone. At distance $d$, the isochrone is shifted in magnitude space by the relevant distance modulus, and the corresponding set of $I$ weights is noted $\mathcal{W}_d = \{ w_i \}_{i=1}^I$. One can repeat this procedure for $D$ different distances. We note $\mathcal{W} = \{ \mathcal{W}_d \}_{d=1}^D$ as the set of sets of $I$ weights.

\subsection{Spatial enhancement}
\label{sub_sec:spatial_enhancement}
The algorithm gnomonically projects the stars into a tangential plane. They are then spatially binned in $N \times M$ pixels of indices ($n,\, m$) and weighted by each $\{ \mathcal{W}_d \}_{d=1}^D$, resulting in a set of weighted density maps, $\{ \mathcal{M}_d \}_{d=1}^D$.

It is essential to account for irregularities and holes in the footprint to avoid spurious detections of stellar overdensities related to an imperfect assessment of the field contamination. Momentarily disregarding the weights, the algorithm groups pixels together (typically 8 by 8), effectively performing a second larger binning. If none of the pixels from a group contains a star, all underlying pixels are classified as empty ($=0$). If at least one pixel in the group contains a star, all are marked as non-empty ($=1$). This procedure is used to produce an approximated footprint, which we denote as $\mathcal{F}$.

To enhance overdensities of varying angular scales, each $\{ \mathcal{M}_d \}_{d=1}^D$ map is convolved with $K$ 2D Gaussian kernels $\mathcal{G} = \{ g_k\}_{k=1}^K$. The resulting smoothed maps are
\begin{eqnarray}
	\mathcal{M}_{(d,\, k)} = \mathcal{M}_d \ast g_k,
	\label{eq:smoothed_maps}
\end{eqnarray}  
with the asterisk ($\ast$) representing the convolution operator\footnote{All convolutions were computed using fast Fourier transform.}. The complete set is denoted as $\mathcal{M} = \{ \mathcal{M}_{(d, \, k)} \}_{d, \, k=1}^{D, \, K} $. 

\subsection{Significance maps}
\label{sub_sec:significance_maps}
Comparing the $D \times K$ maps in $\mathcal{M}$ for different distances and convolution kernels requires assessing the significance of a pixel in the various maps by quantifying the deviation of its value from those of neighbouring pixels. In the following, let us first consider only one of the $\mathcal{M}$ maps. For pixel ($n,\, m$), we estimate the shape ($\alpha$) and scale ($\beta$) parameters of a gamma distribution fitted to the values of surrounding pixels contained within an annular region of constant size, irrespective of the kernel size. We have
\begin{eqnarray}
	\alpha^{(n,\, m)} = \frac{(\,\mu_{\mathrm{ann}}^{(n, \,m)})^2}{v_{\mathrm{ann}}^{(n, \,m)}} \hspace{5mm} \mathrm{and} \hspace{5mm} \beta^{(n,\, m)} = \frac{v_{\mathrm{ann}}^{(n, \,m)}}{\mu_{\mathrm{ann}}^{(n, \,m)}},
	\label{eq:gamma_parameters}
\end{eqnarray}  
where $\mu_{\mathrm{ann}}^{(n, \,m)}$ and $v_{\mathrm{ann}}^{(n, \,m)}$ represent the mean and variance of the pixels within the annulus centred on the pixel ($n,m$) . We set the annulus kernel $a(r_{\mathrm{in}}, r_{\mathrm{out}})$ to 1 for $r_{\mathrm{in}} < r < r_{\mathrm{out}}$ and 0 otherwise, where $r_{\mathrm{in}}$ and $r_{\mathrm{out}}$ are the inner and outer radii of the annulus, respectively. By convolving the $\mathcal{M}_{(d, \, k)}$ map with $a(r_{\mathrm{in}}, r_{\mathrm{out}})$, one can determine the means and variances of all pixels at once using
\begin{eqnarray}
	\mu_{\mathrm{ann}} = \frac{\mathcal{M}_{(d, \, k)} \ast a(r_{\mathrm{in}}, r_{\mathrm{out}})}{\mathcal{F}\ast a(r_{\mathrm{in}}, r_{\mathrm{out}})}
	\label{eq:mean}
\end{eqnarray}  
\begin{eqnarray}
	v_{\mathrm{ann}} = \frac{\mathcal{M}_{(d, \, k)}^{\,2} \ast a(r_{\mathrm{in}}, r_{\mathrm{out}})}{\mathcal{F}\ast a(r_{\mathrm{in}}, r_{\mathrm{out}})} - \frac{\big(\mathcal{M}_{(d, \, k)} \ast a(r_{\mathrm{in}}, r_{\mathrm{out}})\big)^2}{\mathcal{F}\ast a(r_{\mathrm{in}}, r_{\mathrm{out}})},
	\label{eq:variance}
\end{eqnarray}  
where the denominators include the approximation of the footprint $\mathcal{F}$ and act as a renormalisation factor. With the parameters estimated in Eq.~\ref{eq:gamma_parameters}, we evaluate, for each pixel of the $\mathcal{M}_{(d, \, k)}$ map, the value of the corresponding CDF and use it as an assessment of the significance of a detection. We denote $\mathcal{S}_{(d, \, k)}$ the significance map for a given distance $d$ and kernel $k$, and we define its value for pixel ($n, \, m$) as
\begin{eqnarray}
	\mathcal{S}_{(d, \, k)}^{(n, \,m)} = -\ln(1 - \mathrm{CDF}_{(d, \, k)}^{(n, \,m)}).
	\label{eq:significance}
\end{eqnarray}  

Repeating this procedure for each map in $\mathcal{M}$ yields $\mathcal{S} = \{ \mathcal{S}_{(d, \, k)} \}_{d, \, k=1}^{D, \, K}$, the set of significance maps which can be directly compared with one another. Finally, we construct our final significance map, $\mathcal{S}_\mathrm{max}$, as the map for which the value in pixel $(n, \, m)$ is defined as the maximum value of $(n, \, m)$ for the different pixels across all maps of $\mathcal{S}$.

Peaks are found within this map, and their values are associated with overdensities -- namely, an ensemble of neighbouring pixels exceeding a specified threshold. As the algorithm also tracks for each pixel the distance shift ($d$) and kernel index ($k$) that maximised its significance, \textsc{erose} returns these values as estimates of the overdensity's distance and angular size. We note that the significance values returned by \textsc{erose} range from 0 to 36.74, with the upper limit being purely numerical and corresponding to the double-precision floating-point arithmetic. 


\section{Simulated LSST data and injection of artificial satellites}
\label{sec:data}
\subsection{The LSST DC2 simulation}
\label{sub_sec:what_dc2}
The analysis presented in this work used the static component of the LSST Data Challenge 2 (DC2) simulation \citep{Abolfathi_2021}, which was carried out by the LSST Dark Energy Science Collaboration (LSST DESC) to provide a test bed for source detection algorithms, photometric measurements, and scientific analyses. The DC2 simulation covers $\sim$300 deg$^2$ of the wide-fast-deep (WFD) area that will be surveyed by LSST, at relatively high galactic latitude to avoid significant contamination from the Milky Way. It has an extragalactic component based on the \textsc{cosmoDC2} simulation \citep{Korytov_2019} and a local component based on \textsc{galfast}, which generates mock catalogues of Milky Way stars with realistic, empirically driven properties (e.g. see \citet{Juric_2008} for the stellar density laws and \citet{Ivezic_2008} for the metallicity distribution). The DC2 simulation emulates the first five years of observations, corresponding to the planned sixth Rubin data release, DR6, accounting for the expected cadence, seeing, and noise. It uses version 19.0.0 of the LSST science pipelines code for image processing \citep[see][for a full overview]{Bosch_2018}. 

Two main catalogue products are available in DC2, the \texttt{Truth\_Match\_Table} and the \texttt{Object\_Table}, both of which are available via the LSST DESC data portal\footnote{\url{https://data.lsstdesc.org/}} \citep[more details in][]{Abolfathi_2021_ReleaseNote}. The \texttt{Truth\_Match\_Table} contains the intrinsic properties of all simulated astrophysical sources, including their true magnitudes, positions, and physical classifications (i.e. stars, galaxies, type Ia supernovae). These quantities represent the `ground truth' of the simulation and are unaffected by observational effects. In contrast, the \texttt{Object\_Table} provides the outputs of the LSST science pipeline applied to the simulated images. It contains astrometric quantities, source characterisations and $ugri$ measurements with 5$\sigma$ depths of 25.16, 26.2, 26.03 and 25.33, respectively. By comparing the true input properties of stars (from \texttt{Truth\_Match\_Table}) to their measured properties (from \texttt{Object\_Table}), one can assess the completeness of the `observations' of the simulated data and characterise their photometric uncertainties (more details in Sect.~\ref{sub_sec:uncert_completeness}).

For the purpose of this study, we defined four slightly overlapping LSST fields in the simulation, each $9 \times 9$ deg$^2$ large. These fields are centred on $(\alpha, \delta) = (61.86\pm4.7\deg, -31.7\deg)$ and $(61.86\pm5.22\deg, -40.0\deg)$ and were used for the injection and recovery of artificial dwarf galaxies (see Sect.~\ref{sub_sec:dwarf_generation} and \ref{sub_sec:implementation}, respectively).

\subsection{Catalogue uncertainties and completeness}
\label{sub_sec:uncert_completeness}
\begin{figure}
	\centering
	\includegraphics[width=\hsize]{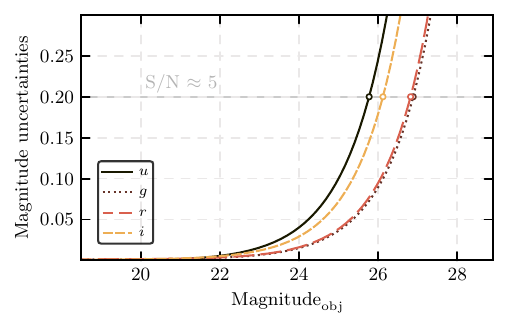}
	\caption{Fitted photometric uncertainties of point-like sources as a function of the LSST measured magnitudes.}
	\label{fig:unc_vs_mag}
\end{figure}

To characterise the photometric uncertainties, we selected stars from the \texttt{Object\_Table} within a 5 deg$^2$ region around $(\alpha, \delta) =(62\deg, -37\deg)$, with a star identified as an object for which \texttt{extendedness == 0}. This parameter distinguishes point-like sources from more extended ones by comparing a point spread function (PSF) model to PSF-convolved galaxy models \citep[see][for a full description of the procedure]{Bosch_2017}. The photometric uncertainties of these stars were then modelled for each of the photometric bands by fitting exponentials as a function of the respective magnitudes \citep[following a procedure similar to][]{Ibata_2007}. These are presented in Fig.~\ref{fig:unc_vs_mag}.

\begin{figure}
	\centering
	\includegraphics[width=\hsize]{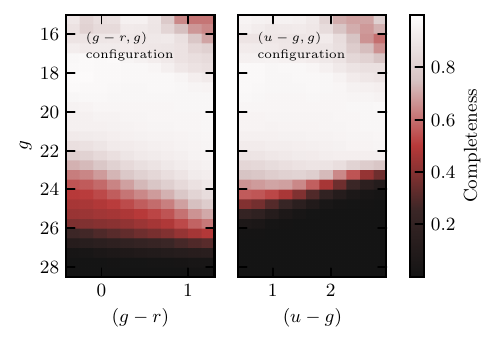}
	\caption{\textit{Left}: Stellar completeness as a function of the input simulation magnitudes and colours for the $(g-r, g)$ photometric configuration. \textit{Right}: Same as above but for the $(u-g, g)$ configuration.}
	\label{fig:completeness}
\end{figure}

As described in \citet{Bosch_2017}, sources are detected independently in each photometric band and then merged into an object catalogue from which each object is assigned a reference photometric band (most often $i$). This band is used both for object characterisation and to perform forced photometry across the other bands, correlating all measurements. For a given combination of $C$ photometric bands (details about each configuration are provided in Sect.~\ref{sub_sec:implementation}), we determined the completeness in a $C$-dimensional (multi-)colour-magnitude space. In practice, we created a $C$-dimensional histogram and compared the number of truth-matched stars, i.e. simulated DC2 stars (\texttt{truth\_type == 2}) with photometric uncertainties $< 0.2$ (S/N $\gtrsim 5$) in all bands of interest and classified by the LSST pipeline as point-like sources (\texttt{extendedness == 0}) to the total number of DC2 stars (\texttt{truth\_type == 2}) in each bin. All histograms for all configurations share the same binning scheme, with bin sizes of 0.23 for $(u-g)$, 0.16 for $(g-r)$, 0.08 for $(r-i)$, and 0.5 for $g$. These values were chosen such that, even for the most demanding configuration, i.e. the 4D ($u-g, \, g-r, \, r-i, \, g$) multi-colour-magnitude space containing the largest number of bins, $\gtrsim$85\% of those contain at least ten stars. To further reduce statistical fluctuations, we smoothed the histogram with a top-hat kernel averaging the values of all direct neighbouring pixels. Fig.~\ref{fig:completeness} displays our completeness estimates for the $(g-r, g)$ and $(u-g, g)$ photometric configurations, since higher-dimensional configurations are impractical to visualise. Both panels exhibit a decline in stellar completeness with increasing magnitude, reflecting the expected lower performances of source detection and stellar classification for progressively fainter objects. The trend observed in the $(g-r, g)$ configuration is mainly driven by stellar classification, producing a smooth magnitude-dependent transition. In contrast, the sharper transition in ($u-g, g$) is mainly due to source detection limits, as the $u$ band’s shallower depth (relative to redder bands) causes our strict photometric uncertainty criteria to act as a hard magnitude cutoff before stellar classification begins to degrade. Assuming those stellar completeness, one can generate DC2-consistent dwarf galaxies (see more details in Sect.~\ref{sub_sec:dwarf_generation}).

\subsection{Dwarf galaxy properties}
\label{sub_sec:dwarf_properties}
The recovery fraction of satellites is assessed in a 3D parameter space defined by the $V$-band absolute magnitude, $M_V$, the half-light radius, $r_h$, and the heliocentric distance, $d$. In practice, we created six logarithmically spaced distance bins in which the satellite distances were linearly drawn. We further binned the $(r_h, M_V)$ space into 600 pixels and drew ten satellites in each, bringing the total number of satellites generated per distance bin to 6000, and to 36\,000 across all distances. 

To be more realistic, additional properties were drawn for the artificial dwarf galaxies, namely ellipticity ($e$), position angle ($\theta_{\mathrm{pos}}$), age ($\tau$), and metallicity ([Fe/H]). The ellipticities were sampled from the smoothed distribution of Local Group dwarf galaxies, derived from the Local Volume Database of \citet{Pace_2025}. Age was fixed to 13 Gyr, as recent studies have found that faint systems such as those we are interested in were often quenched $\sim$13 Gyr ago by supernova feedback and reionisation \citep[e.g.][]{Brown_2014, Gallart_2021, Sacchi_2021, Durbin_2025}. The generation of our mock dwarf galaxies (further detailed in Sect.~\ref{sub_sec:dwarf_generation}) uses a set of luminosity functions and isochrones spanning different metallicities, with a metallicity step of $\Delta$[Fe/H] $= 0.05$, all downloaded from the \textsc{PARSEC} library \citep{Bressan_2012, Marigo_2017}. Using the metallicity-magnitude relation from \citet{Kirby_2013}, $[\mathrm{Fe/H}]=-0.12M_V - 2.91$, fitted using the Milky Way's dwarf spheroidal satellite galaxies, we derived a metallicity for each sampled $M_V$. This allows each dwarf galaxy to be associated with one isochrone and one luminosity function. We note that \textsc{PARSEC} has a metallicity floor at $[\mathrm{Fe/H}]\sim-2.2$ and any metallicity below this threshold was raised to this limit\footnote{Although this truncation arises from purely technical considerations, it is also a reasonable approximation of the luminosity--metallicity relation determined by \citet{Fu_2023}, who show that the metallicities of the faintest systems plateau at low values. Additionally, isochrones are typically very similar in the low-metallicity regime.}. We summarise the properties of our artificial satellites in Table~\ref{tab:dwarf_properties}.

\begin{table}[]
\caption{Properties of the simulated satellites.}
\label{tab:dwarf_properties}
\renewcommand{\arraystretch}{1.2}
\begin{tabular}{lccc}
\hline
\hline  

\multicolumn{1}{c}{Parameter}          & Range                  & Unit & Sampling \\ \hline
Absolute magnitude ($M_V$)           & {[}$-8, 2${]}         & -        & linear      \\
Half-light radius ($r_{\rm h}$)            & {[}$1, 10^3${]}   & pc     & log$_{10}$      \\
Heliocentric distance ($d$)                & {[}$10, 10^3${]} & kpc  & linear      \\
Ellipticity ($e$)                                      & {[}0, 0.8{]}            & -        & linear   \\
Position angle ($\theta_{\rm pos}$) & {[}0, 180{]}           & deg  & linear   \\ \hline
Age ($\tau$)                                          & 13                         & Gyr   & choice     \\
Metallicity ([Fe/H])                               & {[}$-2.2, -1.9${]}  & -        & linear     \\ \hline
\end{tabular}

\end{table}

\subsection{Dwarf galaxy generation}
\label{sub_sec:dwarf_generation}
Our dwarf galaxies consist of an ensemble of individually resolved stars with $(\alpha,\delta)$ positions and photometric information in each of the LSST $ugri$ bands. In the following, we describe how those were generated to be realistic.

\begin{figure}
	\centering
	\includegraphics[width=\hsize]{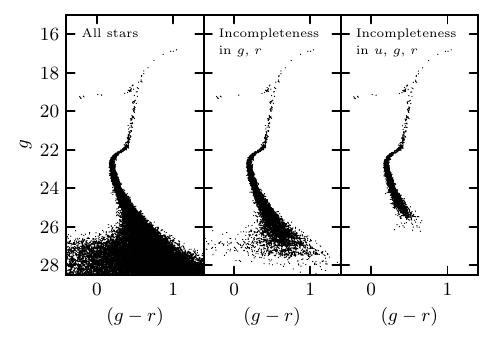}
	\caption{CMD of a dwarf galaxy with an apparent magnitude $M_V = -5.3$, located at a distance of $55~\mathrm{kpc}$. \textit{Left}: CMD of the generated population with a scatter accounting for both the intrinsic dispersion of the population and the photometric uncertainties. \textit{Middle}: Same as in the left panel but also accounting for the combined incompleteness in the $g$ and $r$ bands. \textit{Right}: Same as in the left and middle panels but further accounting for incompleteness in the $u$ band.}
	\label{fig:cmds}
\end{figure}

For a dwarf galaxy with a given metallicity, we first generated its photometry by drawing stars under the corresponding \textsc{PARSEC} luminosity function (using the Oct. 2017 LSST photometric system, made for the DC2 simulation) until the target $M_g$ was reached\footnote{$M_g$ magnitudes were converted to $M_V$ using $M_g = 0.99M_V + 0.17$. These were derived from artificial stellar populations of varying total luminosities generated using PARSEC.}. The remaining $u$, $r$, and $i$ absolute magnitudes were determined from different isochrone colour combinations. We converted the absolute magnitudes to apparent magnitudes using the distance modulus (distance $d$), and added the scatter due to photometric uncertainties (determined in Sect.~\ref{sub_sec:uncert_completeness}), which we inflated with an additional intrinsic population scatter of 0.01 mag. The left panel of Fig.~\ref{fig:cmds} shows the resulting CMD of a generated mock satellite placed at a heliocentric distance of $55~\mathrm{kpc}$. 

Spatially, our galaxies follow an exponential profile:
\begin{eqnarray}
	\Sigma(r) = \frac{1.68^2}{2\pi r_h^2 (1-e)}\exp \Bigg(-1.68 \frac{r}{r_h} \Bigg) ,
	\label{eq:profile}
\end{eqnarray}    
where $r = \sqrt{\big(\xi'/(1-e)\big)^2 + \eta'^2}$ is the radial coordinate in a polar coordinate system. Here, the $\xi'$-axis is scaled by the circularity ($1-e$), and the $\eta'$-axis remains the same. We randomly drew ($\xi'/(1-e), \, \eta'$) star positions from its related probability density function,
\begin{eqnarray}
	p(r) = 2 \pi r \Sigma(r).
	\label{eq:pdf}
\end{eqnarray}    
With a change of variable, one can show from Eq.~\ref{eq:pdf} that $r$ follows a gamma-distribution with a shape parameter $\alpha = 2$ and a rate parameter $\lambda = 1.68/r_h$. Using the corresponding inverse CDF, we obtained the radius ($r_\mathrm{max}$) enclosing 90$\%$ of the stars ($\sim 2.3 r_h$). The final $(\xi, \, \eta)$ positions of the stars were computed using the coordinate transformation:
\begin{eqnarray}
	\begin{pmatrix} \xi \\ \eta \\ \end{pmatrix} = \mathbf{R}(\theta_\mathrm{pos}) \begin{pmatrix} \xi'/(1-e) \\ \eta' \\ \end{pmatrix}
	\label{eq:rotation_matrix}
\end{eqnarray}    
where $\mathbf{R}(\theta_\mathrm{pos})$ denotes the standard 2D rotation matrix.

Following a procedure similar to that described in \citet{Zhang_2025}, we randomly injected dwarf galaxies in each of the four LSST fields (defined in Sect.~\ref{sub_sec:what_dc2}) ensuring they were spaced far enough apart to avoid mutual contamination and enable proper background estimations. This distance between the dwarf galaxies is given by
\begin{eqnarray}
	\Delta_\mathrm{separation} = \mathrm{max} [r_\mathrm{max}, r_\mathrm{ann}],
	\label{eq:ang_separation}
\end{eqnarray}  
where $r_\mathrm{ann}$ is the outer radius of the background annulus used by our search algorithm (see Sect.~\ref{sub_sec:significance_maps}). With the centres of the satellites chosen, one can convert the ($\xi$, $\eta$) coordinates of its stars into $(\alpha,\delta)$ and inject those into one of the 7051 resulting fields, which together cover an area of 571\,131 deg$^2$.

\begin{figure*}
	\centering
	\includegraphics[width=\hsize]{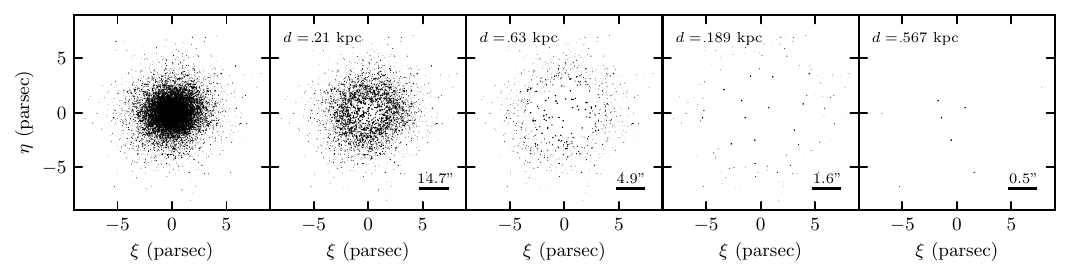}
	\caption{Left-most panel: Real distribution of a dense $r_h = 1.5$ pc satellite with $M_V = -5.7$, assuming all stars could be resolved. Remaining panels, from left to right: Effect of our crowding approximation. As we shift our dwarf galaxy to greater distances, we notice that stars are first lost in the densest region (the core), where the typical distance between two objects is smaller than the expected LSST PSF. For the last distance ($d=567$ kpc), the entire satellite exhibits an angular size comparable to that of the PSF itself, resulting in only few stars being detected. The size of the points corresponds to the magnitude of the stars. We notice that only bright stars are detected within the inner regions of the dwarf galaxy.}
	\label{fig:crowding_effect}
\end{figure*}

As highlighted by \citet{Zhang_2025}, studies aiming to determine dwarf-satellite detection limits \citep[e.g.][]{Koposov_2008, Drlica_Wagner_2020, Tsiane_2025} often overlook the impact of crowding on the detection of stars in dense systems. This oversight can lead to an overestimation of the recovery fraction of the densest and/or most distant satellite systems\footnote{We find that due to the size-luminosity relation of dwarf galaxies, our blending approximation affects $\lesssim 0.1\%$ of simulated Milky Way satellites and, therefore, have a negligible impact on our final estimates. However, blending effects would impact the detectability of more compact objects such as globular clusters.}. To address this issue, we performed a catalogue-level approximation by considering a star to be detected by the pipeline only if it is significantly brighter than its neighbours. Specifically, we required its flux to constitute at least 90$\%$ of the total flux of all stars within a radius of $r_\mathrm{blend}$. This relies on the underlying assumption that, in crowded regions, the PSF blends the faintest sources into a `uniform' extended background on top of which, a star that is sufficiently bright can be detected. For this analysis, we set $r_\mathrm{blend} = 0.425''$. According to the \texttt{baseline\_v5.1} run of the OpSim simulation\footnote{This simulation models Rubin's system throughput and the LSST observing strategy. It is available at \url{https://usdf-maf.slac.stanford.edu/?runId=1}}, the expected median seeing\footnote{The seeing refers to the full width at half maximum (FWHM) of the PSF, where FWHM $\approx2.355\sigma$.} across the DC2 footprint is $\sim$1'' in the worst band ($u$). Fig.~\ref{fig:crowding_effect} illustrates the impact of this approximation for a very dense object ($r_h = 1.5$~pc) shifted to different distances. At the closest distances, most stellar losses occur in the system's very inner region due to high stellar density. As the object is shifted to greater distances, stars in the outer regions are also lost due to crowding and as a result, very few stars remain.

As our closest mock dwarf galaxies lie 10 kpc away, they are located at a distance where 3D modelling of Milky Way dust extinction \citep{Amores_2005} is equivalent to a simple integrated approach. We therefore reddened our stars according to the \citet{Schlegel_1998} integrated dust maps \citep[recalibrated using][]{Schlafly_2011}, ensuring photometric consistency with the DC2 simulation \citep[see Sect.~5.3 of][]{Abolfathi_2021}. Finally, to make our dwarf galaxies fully consistent with the DC2 simulation, we randomly removed stars according to the completeness histograms from Sect.~\ref{sub_sec:uncert_completeness}. The middle and right panels of Fig.~\ref{fig:cmds} show the same CMD as the left panel, this time accounting for the combined incompleteness of using $gr$ and $ugr$, respectively.



\section{Expected detection limits for dwarf galaxies in the LSST}
\label{sec:results}
\subsection{Implementation and performance}
\label{sub_sec:implementation}
We ran \textsc{erose} on each of the 7051 fields, where both the measured star--galaxy separation and the injection of mock dwarf galaxies were already performed (for more details, see Sects~\ref{sub_sec:what_dc2} and \ref{sub_sec:dwarf_generation}, respectively). Such a field contains between 250\,000 to 3\,500\,000 stars, a number that varies due to the completeness reached by each of the six different photometric configurations tested in this study. Each configuration (hereafter referred to by their designations listed in Table~\ref{tab:configuration_summary}) explores a different (multi-)colour-magnitude space, offering distinct insights into the properties of the stars. In the following, we provide the motivations behind choosing each space.
\begin{itemize}
	\item $(g, r)$: This configuration was adopted in most studies \citep[e.g.][]{Belokurov_2007, Koposov_2008, Drlica_Wagner_2015, Tsiane_2025} as its bands are typically the deepest. Moreover, the $(g-r)$ colour provides an efficient way to separate metal-poor from metal-rich stars, in particular, for the main-sequence turn-off.
	
	\item $(r, i)$: This is another widely used combination, particularly in surveys where the $r$ and $i$ bands are the deepest, or $g$ is unavailable \citep[e.g.][]{Laevens_2015, Smith_2025}. It efficiently removes contamination from foreground Milky Way red main-sequence stars.
	
	\item $(g, r, i)$: As discussed in Sect.~\ref{sub_sec:isochrone_selection}, \textsc{erose} assigns a weight for each star corresponding to its probability of belonging to a stellar isochrone in any (multi)-colour-magnitude space. Here, we combined the $(g-r)$ and $(r-i)$ configurations, further improving our ability to suppress contamination.
	
	\item $(u, g)$: As most dwarf galaxies exhibit metallicities below $-$1.5, including the metal-sensitive $u$ band improves our ability to isolate their metal-poor member stars from metal-richer Milky Way disc stars. The ($u, g$) colour also helps separate stars from background galaxies. Together with our strict 0.2 cut in magnitude uncertainty (which acts as a magnitude cutoff removing faint objects poorly classified by the \texttt{extendedness} parameter) our selection yields a stellar sample with a purity of $\sim 98\%$, a significant improvement over that ($\sim 30\%$) achieved using only $(g, r)$.
	
	\item $(u, g, r)$: The $(u-g)$ and $(g-r)$ colour combination was already used in many studies to infer the photometric metallicities of stars \citep[e.g.][]{Ivezic_2008, Ibata_2017}.
		
	\item $(u, g, r, i)$: This configuration combines all aforementioned spaces.
\end{itemize}
\begin{table*}[]
\caption{Photometric configurations, including their designations, derived results, and associated properties.}
\begin{center}
\renewcommand{\arraystretch}{1.5}
\setlength{\tabcolsep}{15pt}
\begin{tabular}{c c c c >{\centering\arraybackslash}p{1.1cm} >{\centering\arraybackslash}p{1.1cm} }
\hline \hline \rule{0pt}{2.2ex}
	(multi-)colour-magnitude space & Designation & Threshold$^{\mathrm{(a)}}$ & Area$^{\mathrm{(b)}}$ & \multicolumn{2}{c}{Recovered satellites$^{\mathrm{(c)}}$} \\ [-0.5ex]
	 & & & (deg$^2$) & ($N$) & (\%)\\ \hline  
	
	($g-r$, $g$) & ($g, r$) & 15.90 & 18,635 & $82^{+17}_{-15}$ & $62.5^{+6.6}_{-6.5}$ \\
	
	($r-i$, $r$) & ($r, i$) & 17.67 & 18,669 & $72^{+15}_{-13}$ & $54.7^{+6.8}_{-6.8}$ \\
	
	($g-r$, $r-i$, $g$) & ($g, r, i$) & 20.29  & 18,635 & $83^{+18}_{-15}$ & $63.5^{+6.4}_{-6.5}$ \\
	
	($u-g$, $g$) & ($u, g$) & 14.95 & 17,197 & $79^{+17}_{-15}$ & $64.1^{+6.3}_{-6.0}$ \\
	
	($g-r$, $u-g$, $g$) & ($u, g, r$) & 18.59 & 17,196 & $75^{+16}_{-14}$ & $61.2^{+6.4}_{-6.4}$ \\
	
	($g-r$, $u-g$, $r-i$, $g$) & ($u, g, r, i$) & 20.61 & 17,196 & $75^{+16}_{-14}$ & $60.8^{+6.5}_{-6.4}$ \\[+0.3ex]\hline 
	
\end{tabular}
\tablefoot{$^{(a)}$Signal metric returned by \textsc{erose} on a scale from 0 to 36.74, more details in Sect.~\ref{sub_sec:significance_maps}. $^{(b)}$Total LSST footprint area using a given set of photometric bands, see Sect.~\ref{sub_sec:nb_frac_satellites} for more details on this estimate. $^{(c)}$Number and fraction of satellites recovered within their respective footprints both provided as the median, 16$^\mathrm{th}$ and 84$^\mathrm{th}$ quantiles of their corresponding distributions (see Sect.~\ref{sub_sec:nb_frac_satellites}).}
  \label{tab:configuration_summary} 
\end{center}
\end{table*}

Stars were weighted (see Sect.~\ref{sub_sec:isochrone_selection}) according to their consistency with a 13 Gyr old, [Fe/H]=$-$2.19 dex isochrone taken from the \textsc{PARSEC} library \citep{Bressan_2012}. In practice, we first spline-interpolated the isochrone using the \texttt{making\_splines} method from \textsc{erose}, producing $\mathcal{Y}$ (with $J =$ 4970). The mask widths were set to 0.05 (with $\boldsymbol{\Sigma}_j$ therefore dominated by the photometric uncertainties). We probed $D=16$ logarithmically spaced distance shifts spanning from 10 kpc to 1000 kpc. The number of dimensions of the input photometric configuration only has a minor impact on the time required to determine the weights.

The typical angular sizes of Milky Way dwarf galaxies range from $\sim1'$ to $\sim20'$, prompting us to use $g_k = \{0.5',\,1',\,2',\,4',\,8',\,16',\,24'\}$ to enhance the detection-signal of overdensities of such sizes. To estimate the local background as close as possible to the overdensity while maintaining $r_\mathrm{in}$ at least 2$\sigma$ away from the largest $g_k$, we set $r_\mathrm{in} = 60'$ and  $r_\mathrm{out} = 100'$. In the closest distance bin, only a few highly extended satellites (the largest objects in the first two distance bins) would require significantly larger values of $g_k$ and $r_\mathrm{in}$ for detection. Increasing $r_\mathrm{in}$ to such an extent would compromise the locality of our background estimate. Such extended systems would likely be detected through other mapping methods anyway, as was the case for the six known Milky Way satellites larger than 25' \citep[e.g.][]{Grillmair_2009, Torrealba_2019}.

With this setup, determining the weights, computing the 112 $\mathcal{M}_{(s,\, k)}$ maps, and extracting overdensities from the final $\mathcal{S}_\mathrm{max}$ map take 1 to 2 minutes on a 2019 Intel MacBook Pro laptop, depending on the chosen photometric configuration. This means that the algorithm could run a dwarf galaxy search over the entire 18\,000 deg$^2$ area of the expected LSST main-survey footprint \citep{Ivezic_2019} in a matter of hours. Moreover, since each field is independent, the search could easily be carried out in parallel on multiple machines.

\subsection{Recovery fraction maps}
\label{sub_section:recovery_fraction}
\begin{figure}
	\centering
	\includegraphics[width=\hsize]{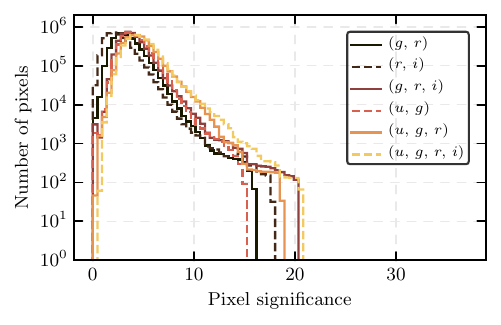}
	\caption{Histograms of the pixel significances returned by the algorithm for the six different photometric configurations, each estimated from running \textsc{erose} on all four DC2 empty fields (presented in Sect.~\ref{sub_sec:what_dc2}). The maximum significance returned by \textsc{erose} is 36.74.}
	\label{fig:histo}
\end{figure}

Determining whether a satellite is detected or not requires defining a detection threshold -- a significance level above which no (or at least very few) false positives are expected. To set this threshold, we first applied our algorithm to each of the four LSST fields without injecting any dwarf galaxy and recorded the resulting pixel significances. This procedure was repeated for the six photometric configurations under consideration. The resulting pixel significance distributions are shown in Fig.~\ref{fig:histo}. All distributions exhibit similar overall behaviour; however, we note that configurations including more bands are shifted towards higher significance values. For each photometric configuration, we set the detection threshold to the maximum significance of the corresponding distribution. These restrictive thresholds are provided in Table~\ref{tab:configuration_summary} and were set to exclude false positives resulting from noise fluctuations. 

\begin{figure*}
	\centering
	\includegraphics[width=\hsize]{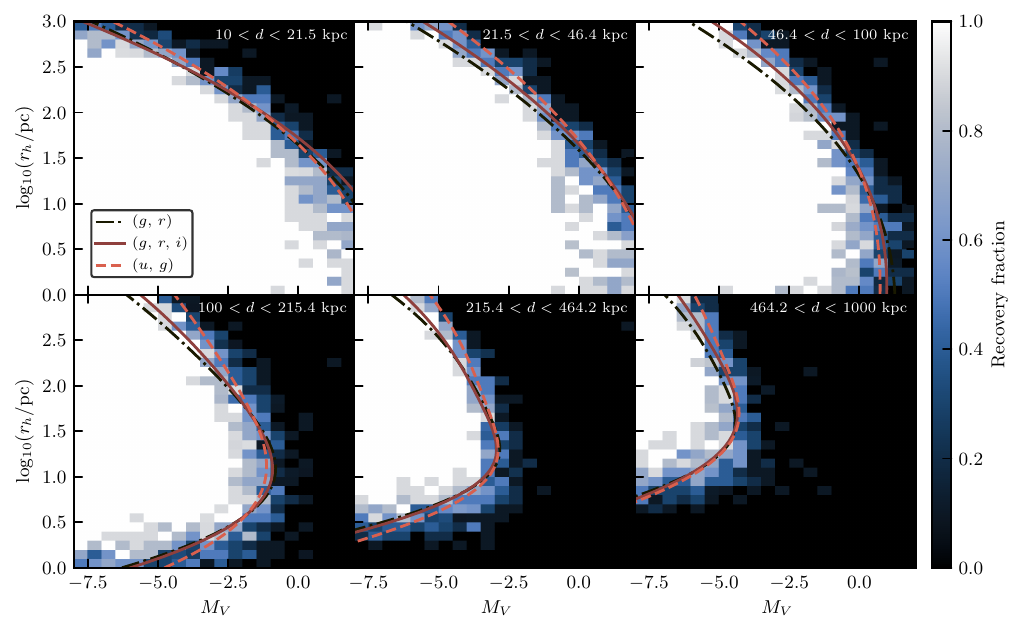}
	\caption{Recovery fraction of the simulated dwarf galaxies in the $(u, g)$ photometric configuration from 100\% (in white) to 0\% (in black). These recovery fractions are provided as a function of the absolute V-band magnitude ($M_V$), physical half-light radius ($r_h$) and heliocentric distance ($d$). We overlay the 50\% recovery fraction limits of the ($g, r$), ($g, r, i$), and ($u, g$) configurations based on our analysis (see more in Fig.~\ref{fig:recovery_fraction_all}).}
	\label{fig:recovery_fraction}
\end{figure*}

Fig.~\ref{fig:recovery_fraction} presents the recovery fraction obtained using the $(u, g)$ photometric configuration. For dwarf galaxies located between 10 and $21.5~\mathrm{kpc}$ (top-left panel), we find that only the most extended and/or least-luminous objects fail to be detected by the algorithm. As distance increases, progressively brighter objects become undetected because their member stars are now too faint. Beyond $100~\mathrm{kpc}$, the recovery fraction of the densest objects also rapidly drops due to crowding, a feature also observed by \citet{Zhang_2025}\footnote{\citet{Zhang_2025} carried out an analysis injecting dwarf galaxies in images, therefore fully simulating crowding effects. The similar trends observed in our results suggest that our crowding approximation introduced in Sect.~\ref{sub_sec:dwarf_generation} is adequate.}. Finally, we also see that the adopted threshold effectively prevents false positives, as shown by the clear transition between fully recovered and fully `unrecovered' galaxies.

This transition is often summarised as a curve, aiming at representing the 50\% recovery fraction limit. Instead, we used an analytical 2D model (an eight parameters 2D sigmoid), which captures the smooth transition between the two regimes. Appendix~\ref{appendix_sec:recovery_fraction_model} includes details about the model itself and shows how it reproduces the data. The best parameters resulting from the inference process for each distance bin and photometric configuration can be found in Tables~\ref{tab:inferred_params_16} through \ref{tab:inferred_params_732}.

For clarity, Fig.~\ref{fig:recovery_fraction} displays only the inferred 50\% recovery fraction curves (defined in Eq.~\ref{eq:blended_parabolas} as a by-product of our full 2D model) for the ($g, r$), ($g, r, i$), and ($u, g$) photometric configurations (the other curves can be found in Fig.~\ref{fig:recovery_fraction_all}). Compared to the ($g, r$) configuration, the ($u, g$) setup shows that the inclusion of the $u$ band -- despite its shallower depth -- slightly improves our results, particularly for the most extended objects. For example, in the case of the largest tested satellites ($r_h=1$~kpc), the detection limits shift by almost two magnitudes in the $46.4$ to $100$~kpc distance bin, as the $u$ band better removes foreground contamination. Similarly, including additional photometric bands also reduces contamination, with the $(g, r, i)$ configuration outperforming both $(g, r)$ and $(r, i)$. However, when the $u$ band is included (e.g. the $(u, g, r)$ and $(u, g, r, i)$ configurations), expanding to higher-dimensional multi-colour-magnitude space slightly worsens our results. This is likely due to reduced stellar completeness and suggests that the $u$ band alone is highly effective at isolating the metal-poor stars in our injected objects.

\subsection{Predictions for LSST dwarf galaxy searches}
\label{sub_sec:nb_frac_satellites}
Another way to interpret these results is to estimate the number of Milky Way satellite galaxies that LSST is expected to discover. To this end, we combined our observational selection functions with the anticipated LSST footprint and an empirical model of the Milky Way satellite population.

\begin{figure}
	\centering
	\includegraphics[width=\hsize]{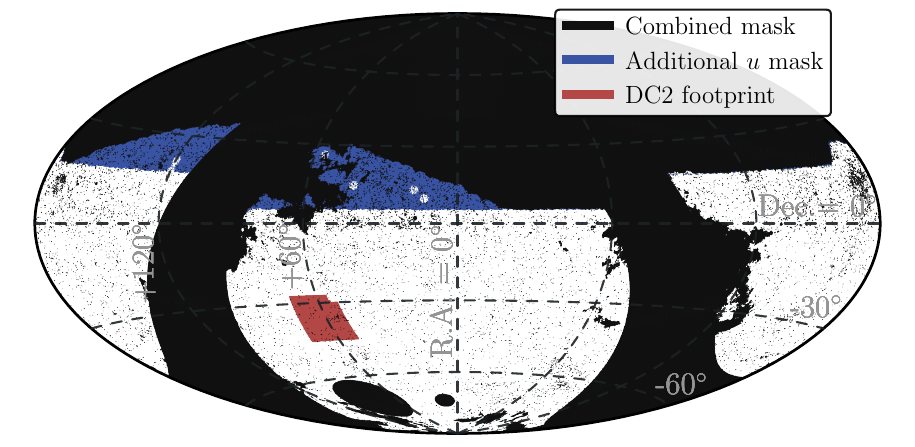}
	\caption{Equatorial Hammer-Aitoff projection of the mask applied in our analysis (black) and the additional area lost when including the $u$ band (blue). For reference, the DC2 footprint is also shown in red.}
	\label{fig:footprint}
\end{figure}

The LSST footprint was obtained from the \texttt{baseline\_v5.1} run of the OpSim simulation, which illustrates the survey strategy over the ten-year of the LSST WFD mission. This simulation accounts for exposure and read-out times in each band, as well as the telescope's slew performance \citep{Jones_2014}. The DC2 field, introduced in Sect.~\ref{sub_sec:what_dc2}, and upon which we build our analysis, has been chosen as a typical `high' galactic latitude field \citep{Abolfathi_2021}. We assumed it is representative of the survey of the Galactic cap and masked out regions at low-galactic latitude ($|b|$ < 15 deg) with high interstellar reddening \citep[$E(B - V) > 0.2$ according to][]{Schlegel_1998}, where the detection of satellite, while not impossible, will likely be severely hampered. Moreover, we also masked regions around known bright stars \citep{Hoffleit_1991}, globular clusters \citep[][2010 edition]{Harris_1996} and nearby semi-resolved galaxies \citep{Nilson_1973, Corwin_2004}. Following a procedure similar to that used by \citet{Drlica_Wagner_2020}, stars and objects without size information were masked within a circular region of 0.1 deg. When size information was available, extended objects were masked within twice their half-light radii, accounting for their ellipticity. Finally, we also masked circular regions of $9\deg$ and $3\deg$ radii around the LMC and SMC, respectively, approximately corresponding to three times their half-light radii \citep{Munoz_2018}. The final combined mask for a given combination of the $g$, $r$, and $i$ band is shown in black in Fig.~\ref{fig:footprint}, as these three bands share nearly identical sky coverage. When the $u$ band is included, the total overlapping area reduces to approximately $1483\,\mathrm{deg}^2$ (blue region in Fig.~\ref{fig:footprint}).

For this study, we needed to assume a model for the distribution of dwarf galaxies around the Milky Way. We used the satellite population model first developed by \citet{Doliva_Dolinsky_2023} for the M31 dwarf galaxy system, updated to the latest knowledge of the Milky Way system \citep{Tan_2025}. The model assumes spherical symmetry and consists of three independent components: 1) the luminosity function of the satellite system, 2) its size-luminosity relation, and 3) the radial distribution of the satellites. This gives a total of six free parameters. In their satellite census study, \citet{Tan_2025} used the combined data from PS1, DES and DELVE to infer the distribution of the Milky Way dwarf galaxy model parameters using a Markov chain Monte Carlo (MCMC) method. They estimate that the Milky Way is surrounded by $265^{+79}_{-47}$ satellites brighter than $M_V = 0$ within 300~kpc, a number consistent with other studies from the literature \citep[e.g.][]{Newton_2018, Nadler_2020, Manwadkar_2022, Ahvazi_2024}. In their paper, \citet{Tan_2025} released a code\footnote{\url{https://github.com/delve-survey/delve_mw_census}} that generates Milky Way satellite populations by first drawing model parameters from the inferred posterior distribution function represented by their MCMC chain, and then produces a realisation of the model based on these parameters. 

\begin{figure}
	\centering
	\includegraphics[width=\hsize]{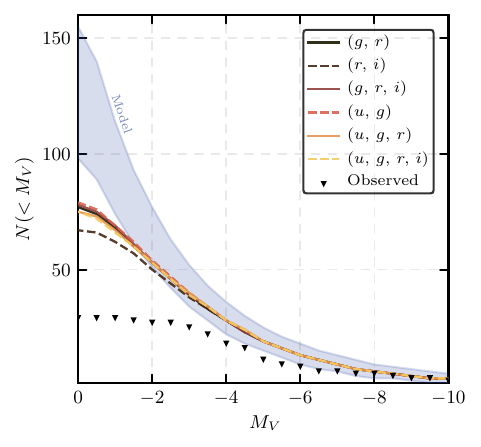}
	\caption{Predicted luminosity functions of the Milky Way satellites, shown as the cumulative number of satellites brighter than a given absolute magnitude ($M_V$) for the different photometric configurations. The empirical model from \citet{Tan_2025} and its 68\% confidence intervals is shown in blue while the black triangles denote the luminosity function of the currently known and confirmed Milky Way dwarf galaxies \citet{Pace_2025}. All luminosity functions are shown for satellites within the LSST WFD survey and covered by all four $ugri$ bands.}
	\label{fig:luminosity_function}
\end{figure}

\begin{figure}
	\centering
	\includegraphics[width=\hsize]{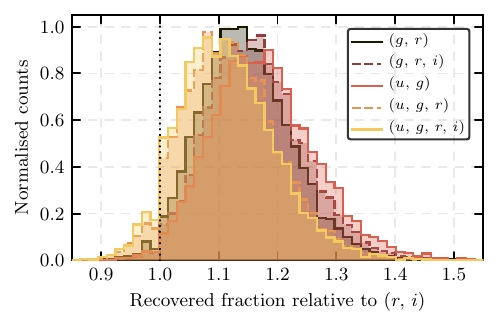}
	\caption{Histograms of the relative number of satellites recovered in the $(g,r)$, $(g,r,i)$, $(u,g)$, $(u,g,r)$, and $(u,g,r,i)$ photometric configurations compared to the $(r, i)$ configuration.}
	\label{fig:distributions}
\end{figure}

We used this code to simulate ten thousand realisations of Milky Way satellite populations. The model's predicted luminosity function contained within the anticipated LSST WFD $ugri$ footprint (with 68\% confidence intervals) is illustrated in blue, in Fig.~\ref{fig:luminosity_function}, as the cumulative number of satellites brighter than a certain absolute magnitude $M_V$. Using our full 2D model (see Appendix~\ref{appendix_sec:recovery_fraction_model}) we were able to determine the detection probability of each satellite depending on their ($M_V$, $r_h$, $d$). To access whether a given satellite was detected, we randomly drew a number between 0 and 1. If its value fell below the satellite's detection probability, it was considered detected; otherwise, it was not. The distributions of the number and fraction of recovered satellites are summarised in Table~\ref{tab:configuration_summary} for each photometric configuration, using their median, 16$^\mathrm{th}$ and 84$^\mathrm{th}$ quantiles.  We find the $(u,g)$ configuration recovers, on average, the largest fraction of input satellites, reaching $64.1^{+6.3}_{-6.0}\%$, compared to $63.5^{+6.4}_{-6.5}\%$ for $(g,r,i)$. Although this difference is not statistically significant, it suggests that the $u$ band can provide additional information and may slightly improve recovery performance despite its shallower depth. However, because of its larger footprint, it is the $(g,r,i)$ configuration that retrieves, on average, the highest number of input satellites, outperforming both the $(g,r)$ and $(r, i)$ configurations. This is further illustrated in Fig.~\ref{fig:distributions} where we compare the number of recovered satellites in each photometric configuration (within the $ugri$ footprint) to the number of recovered satellites using $(r, i)$, i.e. the configuration arguably yielding the lowest recovery performance (see  Figs.~\ref{fig:luminosity_function} \& \ref{fig:recovery_fraction_all}). All configurations almost systematically outperform the $(r, i)$ configuration, recovering on average 10 to 20\% more satellites. We also recover the trend expected from Sect.~\ref{sub_section:recovery_fraction}, where the $(u, g, r)$ and ($u, g, r, i$) configurations underperform relative to $(u, g)$, most likely due to a lower level of stellar completeness.

The coloured lines in Fig.~\ref{fig:luminosity_function} represent our predictions of the observed luminosity function of Milky Way satellites (within the area covered by $ugri$) when using different combinations of photometric bands. They are presented as the median number of satellites recovered for each of the ten thousand population realisations. By comparing our predictions to the currently known population of Milky Way satellites (black triangles), we find that LSST will substantially improve our ability to detect satellite galaxies. This improvement can be expected to be greatest for systems fainter than $M_V \approx -2.0$, and should result in the discovery of $\sim 50$ new satellites.

\section{Summary and discussion}
\label{sec:discussion}
In this study we introduced a new publicly available algorithm, \textsc{erose}, that enables very efficient automated searches of stellar overdensities in any input photometric catalogue, using any combination of photometric bands. We injected a total of 36\,000 mock dwarf galaxies of various properties within the LSST DC2 simulated sky, a simulation designed to represent the first five years of LSST observations. Running \textsc{erose} using six different combinations of photometric bands, we determined the recovery fraction of Milky Way satellites for each as a function of distance, total magnitude, and size (see Fig.~\ref{fig:recovery_fraction}). These comparisons were made possible by the high efficiency of the \textsc{erose} algorithm, expected to run a dwarf galaxy search over the full LSST sky in a matter of hours on a current laptop. 

Combining these detection limits with the LSST expected footprints (Fig.~\ref{fig:footprint}) and one of the latest empirical model of Milky Way satellite population, we were able to estimate the number of dwarf galaxies that LSST might detect. We find that the $(g,r,i)$ configuration recovers, on average, the greatest number of satellites, outperforming the more commonly used $(g, r)$ and $(r,i)$ configurations. This demonstrates the value of adding more photometric bands to more effectively isolate dwarf galaxy member stars from contaminating sources. Configurations including the $u$ band recover fewer satellites because their survey footprint is smaller by 1483 deg$^2$. When considering the fraction of injected satellites recovered, the $(u,g)$ configuration achieves the highest value of recovered satellites, slightly above the $(g,r,i)$ configuration. In contrast to the previous comparison, configurations including the $u$ band together with additional photometric bands, i.e. $(u,g,r)$ and $(u,g,r,i)$, detect fewer satellites most likely due to lower levels of stellar completeness. This underscores the critical role of the $u$ band for improving satellite recovery. Moreover, we see from Fig.~\ref{fig:recovery_fraction} that using the $u$ band could be particularly beneficial for detecting extended objects, for which distinguishing metal-poor member stars from the metal-rich foreground becomes increasingly important.

Given the $\sim 30$ confirmed Milky Way dwarf galaxies currently in the LSST footprint, we expect LSST to more than double this population, with approximately $\sim 50$ new potential discoveries. These new satellites will be found primarily at the faint end of the luminosity function (as shown in Fig.~\ref{fig:luminosity_function}), where their properties provide some of the strongest constraints on dark matter models. One example is Aquarius IV, an ultra-faint Milky Way satellite ($M_V=-1.9^{+0.6}_{-1.0}$), discovered by \citet{Cerny_2026} in the Rubin Early Data Preview 2. In addition to Milky Way satellites, our results also show that systems with absolute magnitudes as faint as $M_V=-4.0$ could be detected up to the edge of the Local Group at 1 Mpc.

Our results are in agreement with those of \citet{Tsiane_2025}, who conducted a similar study using the \textsc{simple} algorithm \citep{Bechtol_2015, Drlica_Wagner_2020}\footnote{\url{https://github.com/sidneymau/simple_adl/tree/kb}} only considering the $g$ and $r$ photometric bands. Their study demonstrates the importance of star--galaxy separation, as their different estimates using either ideal, measured or measured-corrected star--galaxy separation yields $89\pm20$, $83\pm18$ or $67\pm14$ recovered satellites, respectively. Because of our strict thresholds (discussed in Sect.~\ref{sub_section:recovery_fraction}), our rate of false detections is negligible and our results should therefore be compared to their measured-corrected results, which yield a false positive rate of 2.3\%, against 22.4\% for their measured estimate. In this case, they recover $67\pm14$ satellites, a number consistent within 1$\sigma$ with our estimated recovery of $82^{+17}_{-15}$ satellites when using the same $(g, r)$ photometric configuration. This difference is most likely driven by the underlying Milky Way satellite population models. Specifically, we adopted the model of \citet{Tan_2025}, which predicts a satellite population of $265^{+79}_{-47}$, whereas the model of \citet{Nadler_2020}, used by \citet{Tsiane_2025}, predicts a smaller population of $220^{+50}_{-50}$ satellites. These differences could also arise from differences in the performance of \textsc{erose} and \textsc{simple}; however, a proper comparison of the two algorithms lies beyond the scope of the present study.

Altogether, these results highlight that combining the use of the $u$ band or a multi-colour-magnitude weighting of the stars with an improved star--galaxy separation could reveal many satellites that are still undetectable. The expected coverage and exceptional depth of the forecasted LSST $ugri$ bands could be used in tandem with data from \textit{Euclid} or the Nancy Grace Roman Space Telescope, as their space-based imaging capabilities would significantly improve star--galaxy separation, especially at very faint magnitudes. The upcoming Chinese Space Station Telescope (CSST; \citealt{CSST_2025}), a survey expected to combine high-quality space-based star--galaxy separation along with deep $ugri$ photometry, should also be very efficient at detecting resolved dwarf galaxies. The work of \citet{Qu_2023} estimated that CSST would achieve recovery fractions comparable to those presented in this study, which could be further improved by the including the $u$ band. Given these findings, we strongly advocate for the use of well calibrated $u$-band photometry in the search of dwarf galaxies, as its shallower depth is more than compensated by its ability to isolate the (very) metal-poor stars that belong to dwarf galaxies from the Milky Way foreground contamination.



\begin{acknowledgements}
      We thank the anonymous referee for their insightful comments and constructive suggestions, which have helped us improve this manuscript. Co-funded by the European Union (Widening Participation, ExGal-Twin, GA 101158446). ES acknowledges funding through VIDI grant "Pushing Galactic Archaeology to its limits" (with project number VI.Vidi.193.093) which is funded by the Dutch Research Council (NWO). This research has been partially funded from a Spinoza award by NWO (SPI 78-411). This work has received funding from the European Research Council (ERC) under the Horizon Europe research and innovation programme (Acronym: EARLYMW, Grant number: 101170507). The analysis has benefited from the use of the following packages: \textsc{numpy} \citep{Harris_2020}, \textsc{scipy} \citep{Virtanen_2020}, \textsc{pandas} \citep{McKinney_2010}, \textsc{matplotlib} \citep{Hunter_2007}, \textsc{astropy} \citep{The_Astropy_Collaboration_2022}, \textsc{numpyro} \citep{Phan_2019}, \textsc{gcrcatalogs} \citep{Mao_2018}.
\end{acknowledgements}

\bibliographystyle{aa}
\bibliography{refs}

\begin{appendix}
    \section{Recovery fraction model}
\label{appendix_sec:recovery_fraction_model}
We start by building an analytic model to represent the 50\% recovery fraction limit. In light of Fig.~\ref{fig:recovery_fraction}, we choose to model it as the sum of two parabolas blended together using a sigmoid:
\begin{eqnarray}
	h(x) & = & (b-a)(x - x_{\mathrm{shift}})^2 \times \frac{1}{1 + \mathrm{e}^{\alpha (x - x_{\mathrm{shift}})}} \nonumber \\
	& & + \, a(x - x_{\mathrm{shift}})^2 + y_{\mathrm{shift}},
	\label{eq:blended_parabolas}
\end{eqnarray}
where ($\log_{10}(r_h)$, $M_V$) have been replaced by ($x$, $y$) for readability. The parameters $a$ and $b$ are the coefficients of both parabolas, $\alpha$ controls how abruptly we go from one to the other, and $x_{\mathrm{shift}}$ and $y_{\mathrm{shift}}$ are shifting the whole curve along the $x$ and $y$ axis, respectively (i.e. $\log_{10}(r_h)$ and $M_V$).

We now make the model more complex in order to fully capture the transition in our 2D space. In Eq.~\ref{eq:blended_parabolas}, $(x - x_{\mathrm{shift}})$ can be interpreted as a measurement of the distance from $x$ to the sigmoid's `turning point' $x_{\mathrm{shift}}$, where the transition occurs. By analogy, a 2D sigmoid defined along a curve requires knowing the distance of any point to that curve. Taking $h(x)$ to be that curve, this distance can be defined as the first-order approximation of the perpendicular distance of $h(x)$ to the point $(x_0, y_0)$
\begin{eqnarray}
	\phi (x_0, y_0) = \frac{y_0 - h(x_0)}{\sqrt{1 + (h'(x_0))^2}}, 
	\label{eq:perp_dist}
\end{eqnarray}
where
\begin{eqnarray}
	h'(x_0) = 2 a x_0 + \frac{2(b-a)x_0}{1 + \mathrm{e}^{-\alpha x_0}} + \frac{\alpha (b-a) x_0^2 \mathrm{e}^{-\alpha x_0}}{(\mathrm{e}^{-\alpha x_0})^2}
	\label{eq:blended_parabolas_derivative}
\end{eqnarray}
is the derivative of $h(x)$ evaluated at $x_0$. With the distance function $\phi$, we can define our 2D sigmoid as
\begin{eqnarray}
	\psi (x, y) = \frac{1}{1 + \mathrm{e}^{-k(x)\phi(x, y)}},
	\label{eq:main_2D_sigmoid}
\end{eqnarray}    
where
\begin{eqnarray}
	k(r_h) = k_1 x^2 + k_2 x + k_3
	\label{eq:transition_polynomial}
\end{eqnarray}
is a polynomial (of parameters $k_1$, $k_2$, and $k_3$) effectively changing the width of the transition as a function of $x$. 

For each of the six maps of each of the six photometric configurations, we infer the eight parameters of the model (Eq.~\ref{eq:main_2D_sigmoid}) using the \textsc{numpyro} library \citep{Phan_2019, Bingham_2019}, employing the No-U-Turn Sampler (NUTS) to perform efficient Hamiltonian Monte Carlo (HMC) inferences. To ensure that the model properly fits the data, we applied visually informed priors to the $M_{V, \mathrm{shift}}$ and $r_{h, \mathrm{shift}}$ parameters, constraining the range within which the bend occurs. 

\begin{figure}
	\centering
	\includegraphics[width=0.75\hsize]{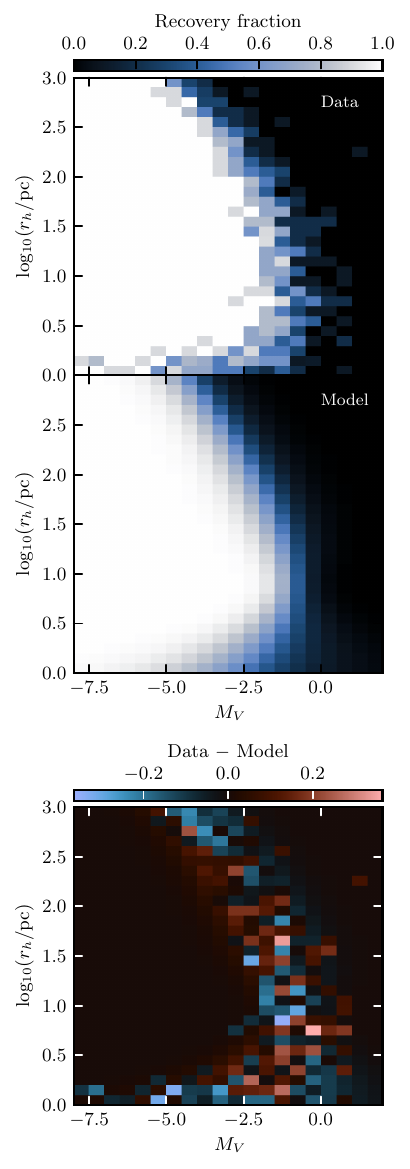}
	\caption{\textit{Top panel}: Recovery fraction of the simulated dwarf galaxies in the $(u, g)$ photometric configuration in the distance bin centred around $158~\mathrm{kpc}$. \textit{Middle panel}: Model resulting from the best-fit parameters inferred via our HMC. \textit{Bottom panel}: Residuals comparing the data to the model.}
	\label{fig:model_vs_data}
\end{figure}

The top panel of Fig.~\ref{fig:model_vs_data} presents the observed recovery fraction for the ($u, g$) configuration in the distance bin spanning from $100$ to $215.4~\mathrm{kpc}$. The middle panel illustrates the corresponding predictions from our analytical model. The bottom panel shows the residuals (data $-$ model) and demonstrates that the model is a good fit to the data: the residuals are predominantly centred around zero, with occasional higher values as a result from statistical fluctuations.

\begin{figure*}
	\centering
	\includegraphics[width=\hsize]{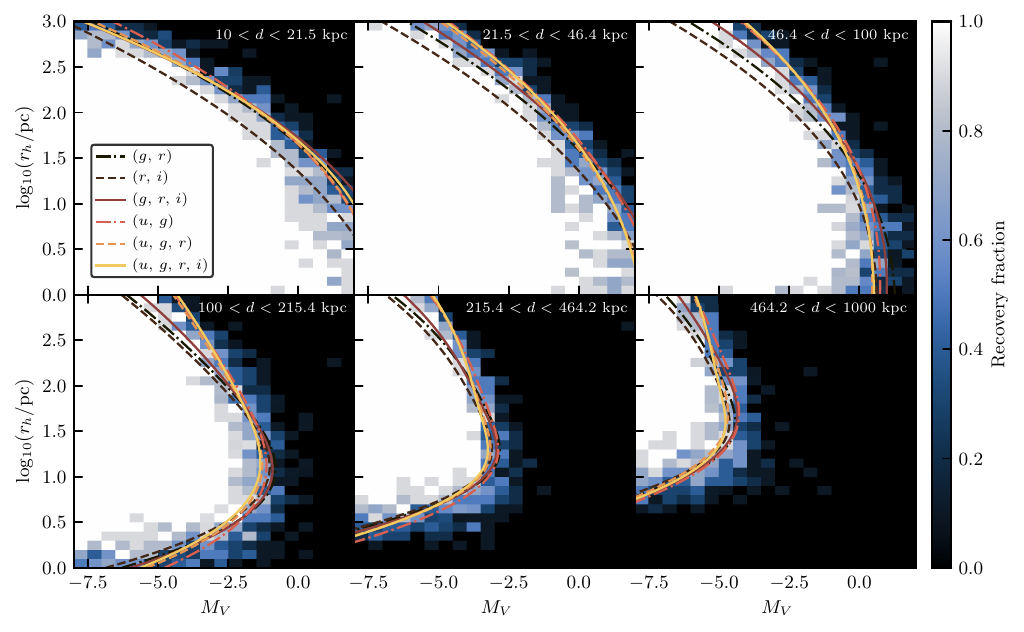}
	\caption{Same as Fig.~\ref{fig:recovery_fraction} but for all the photometric configurations explored in this study.}
	\label{fig:recovery_fraction_all}
\end{figure*}

The eight parameters of the model are inferred for every 2D sigmoids of every distance bins of every photometric configurations. The resulting parameters are all provided in Table~\ref{tab:inferred_params_16} through \ref{tab:inferred_params_732} and are also available in digital form\footnote{\url{https://github.com/samuelrusterucci/erose/tree/main/examples/rusterucci_2026}}. A by-product from our model is the 50\% recovery fraction limit, simply given by Eq.~\ref{eq:blended_parabolas}, shown as the curved lines in Figs.~\ref{fig:recovery_fraction} \& Fig.~\ref{fig:recovery_fraction_all}.
\begin{table*}[]
\caption{Model parameters for distances between 10 and 21.5 kpc.}
\begin{center}
\renewcommand{\arraystretch}{1.1}
\setlength{\tabcolsep}{11pt}
\begin{tabular}{c | c c c c c c c c}
\hline \hline \rule{0pt}{2.2ex}
	Configuration &  $a$ & $b$ & $\alpha$ & $M_{V, \mathrm{shift}}$ & $r_{h, \mathrm{shift}}$ & $k_1$ & $k_2$ & $k_3$ \\ \hline
	$(g,r)$ & $ -0.976 $ & $ 1.748 $ & $ -0.228 $ & $ 3.484 $ & $ -2.45 $ & $ -0.527 $ & $ -0.046 $ & $ 1.523 $ \\
	
	$(r,i)$ & $ 1.228 $ & $ -0.762 $ & $ 0.237 $ & $ 3.806 $ & $ -2.802 $ & $ -0.558 $ & $ -0.039 $ & $ 1.631 $ \\
	
	$(g,r,i)$ & $ 2.183 $ & $ -1.205 $ & $ 0.232 $ & $ 3.341 $ & $ -2.145 $ & $ -0.515 $ & $ -0.06 $ & $ 1.314 $ \\
	
	$(u,g)$ & $ -1.035 $ & $ 1.665 $ & $ -0.216 $ & $ 2.954 $ & $ -2.13 $ & $ -0.396 $ & $ -0.071 $ & $ -0.033 $ \\
	
	$(u,g,r)$ & $ -1.296 $ & $ 2.168 $ & $ -0.231 $ & $ 2.545 $ & $ -1.775 $ & $ -0.484 $ & $ -0.09 $ & $ 0.278 $ \\ 
	
	$(u,g,r,i)$ & $ -1.375 $ & $ 2.343 $ & $ -0.242 $ & $ 2.706 $ & $ -1.691 $ & $ -0.498 $ & $ -0.09 $ & $ 0.458 $ \\ \hline
\end{tabular}
  \label{tab:inferred_params_16} 
\end{center}
\end{table*}

\begin{table*}[]
\caption{Model parameters for distances between 21.5 and 46.4 kpc.}
\begin{center}
\renewcommand{\arraystretch}{1.1}
\setlength{\tabcolsep}{11pt}
\begin{tabular}{c | c c c c c c c c}
\hline \hline \rule{0pt}{2.2ex}
	Configuration &  $a$ & $b$ & $\alpha$ & $M_{V, \mathrm{shift}}$ & $r_{h, \mathrm{shift}}$ & $k_1$ & $k_2$ & $k_3$ \\ \hline
	$(g,r)$ & $ -0.568 $ & $ 0.894 $ & $ -0.323 $ & $ 3.883 $ & $ -2.363 $ & $ -0.545 $ & $ -0.05 $ & $ 1.537 $ \\
	
	$(r,i)$ & $ -0.631 $ & $ 0.822 $ & $ -0.261 $ & $ 3.594 $ & $ -2.495 $ & $ -0.596 $ & $ -0.054 $ & $ 1.89 $ \\
	
	$(g,r,i)$ & $ -0.636 $ & $ 0.872 $ & $ -0.266 $ & $ 3.63 $ & $ -2.235 $ & $ -0.504 $ & $ -0.05 $ & $ 1.163 $ \\
	
	$(u,g)$ & $ -0.58 $ & $ 0.667 $ & $ -0.237 $ & $ 3.432 $ & $ -2.208 $ & $ -0.423 $ & $ -0.054 $ & $ 0.636 $ \\
	
	$(u,g,r)$ & $ -0.741 $ & $ 0.965 $ & $ -0.201 $ & $ 2.612 $ & $ -2.098 $ & $ -0.416 $ & $ -0.069 $ & $ 0.485 $ \\ 
	
	$(u,g,r,i)$ & $ -0.802 $ & $ 1.099 $ & $ -0.199 $ & $ 2.335 $ & $ -1.973 $ & $ -0.412 $ & $ -0.069 $ & $ 0.292 $ \\ \hline
\end{tabular}
  \label{tab:inferred_params_34} 
\end{center}
\end{table*}

\begin{table*}[]
\caption{Model parameters for distances between 46.4 and 100.0 kpc.}
\begin{center}
\renewcommand{\arraystretch}{1.1}
\setlength{\tabcolsep}{11pt}
\begin{tabular}{c | c c c c c c c c}
\hline \hline \rule{0pt}{2.2ex}
	Configuration &  $a$ & $b$ & $\alpha$ & $M_{V, \mathrm{shift}}$ & $r_{h, \mathrm{shift}}$ & $k_1$ & $k_2$ & $k_3$ \\ \hline
	$(g,r)$ & $ 1.615 $ & $ -0.665 $ & $ 0.324 $ & $ 0.781 $ & $ -1.939 $ & $ -0.685 $ & $ -0.09 $ & $ 2.191 $ \\
	
	$(r,i)$ & $ -0.729 $ & $ 1.337 $ & $ -0.479 $ & $ 0.559 $ & $ -0.981 $ & $ -0.957 $ & $ -0.258 $ & $ 0.342 $ \\
	
	$(g,r,i)$ & $ 1.622 $ & $ -0.695 $ & $ 0.31 $ & $ 0.498 $ & $ -1.719 $ & $ -0.619 $ & $ -0.097 $ & $ 1.131 $ \\
	
	$(u,g)$ & $ 1.278 $ & $ -0.578 $ & $ 0.332 $ & $ 0.46 $ & $ -1.473 $ & $ -0.515 $ & $ -0.096 $ & $ 0.277 $ \\
	
	$(u,g,r)$ & $ -0.589 $ & $ 1.264 $ & $ -0.319 $ & $ 0.19 $ & $ -1.41 $ & $ -0.523 $ & $ -0.105 $ & $ 0.067 $ \\ 
	
	$(u,g,r,i)$ & $ -0.711 $ & $ 0.935 $ & $ -0.276 $ & $ 0.523 $ & $ -0.945 $ & $ -0.636 $ & $ -0.191 $ & $ -0.271 $ \\ \hline
\end{tabular}
  \label{tab:inferred_params_73} 
\end{center}
\end{table*}

\begin{table*}[]
\caption{Model parameters for distances between 100.0 and 215.4 kpc.}
\begin{center}
\renewcommand{\arraystretch}{1.1}
\setlength{\tabcolsep}{11pt}
\begin{tabular}{c | c c c c c c c c}
\hline \hline \rule{0pt}{2.2ex}
	Configuration &  $a$ & $b$ & $\alpha$ & $M_{V, \mathrm{shift}}$ & $r_{h, \mathrm{shift}}$ & $k_1$ & $k_2$ & $k_3$ \\ \hline
	$(g,r)$ & $ -5.43 $ & $ -1.145 $ & $ 1.332 $ & $ -0.808 $ & $ 1.089 $ & $ -0.344 $ & $ -0.302 $ & $ -5.573 $ \\
	
	$(r,i)$ & $ -1.129 $ & $ -5.63 $ & $ -1.31 $ & $ -1.018 $ & $ 1.107 $ & $ -0.287 $ & $ -0.242 $ & $ -6.372 $ \\
	
	$(g,r,i)$ & $ -1.086 $ & $ -4.926 $ & $ -1.513 $ & $ -0.914 $ & $ 1.085 $ & $ -0.293 $ & $ -0.258 $ & $ -5.587 $ \\
	
	$(u,g)$ & $ -0.681 $ & $ -3.472 $ & $ -1.241 $ & $ -1.129 $ & $ 1.118 $ & $ -0.253 $ & $ -0.206 $ & $ -4.075 $ \\
	
	$(u,g,r)$ & $ -0.65 $ & $ -3.811 $ & $ -1.291 $ & $ -1.285 $ & $ 1.131 $ & $ -0.205 $ & $ -0.165 $ & $ -4.436 $ \\ 
	
	$(u,g,r,i)$ & $ -3.654 $ & $ -0.674 $ & $ 1.423 $ & $ -1.35 $ & $ 1.161 $ & $ -0.21 $ & $ -0.121 $ & $ -4.267 $ \\ \hline
\end{tabular}
  \label{tab:inferred_params_158} 
\end{center}
\end{table*}

\begin{table*}[]
\caption{Model parameters for distances between 215.4 and 464.2 kpc.}
\begin{center}
\renewcommand{\arraystretch}{1.1}
\setlength{\tabcolsep}{11pt}
\begin{tabular}{c | c c c c c c c c}
\hline \hline \rule{0pt}{2.2ex}
	Configuration &  $a$ & $b$ & $\alpha$ & $M_{V, \mathrm{shift}}$ & $r_{h, \mathrm{shift}}$ & $k_1$ & $k_2$ & $k_3$ \\ \hline
	$(g,r)$ & $ -1.303 $ & $ -7.02 $ & $ -3.012 $ & $ -2.813 $ & $ 1.299 $ & $ -0.367 $ & $ 0.036 $ & $ -6.082 $ \\
	
	$(r,i)$ & $ -1.271 $ & $ -6.92 $ & $ -2.848 $ & $ -3.09 $ & $ 1.299 $ & $ -0.367 $ & $ 0.033 $ & $ -6.102 $ \\
	
	$(g,r,i)$ & $ -1.094 $ & $ -6.528 $ & $ -2.706 $ & $ -2.931 $ & $ 1.299 $ & $ -0.31 $ & $ 0.018 $ & $ -6.462 $ \\
	
	$(u,g)$ & $ -0.798 $ & $ -5.24 $ & $ -2.681 $ & $ -2.873 $ & $ 1.298 $ & $ -0.24 $ & $ 0.065 $ & $ -5.863 $ \\
	
	$(u,g,r)$ & $ -0.706 $ & $ -5.694 $ & $ -3.057 $ & $ -3.145 $ & $ 1.298 $ & $ -0.238 $ & $ 0.111 $ & $ -5.967 $ \\ 
	
	$(u,g,r,i)$ & $ -0.664 $ & $ -5.475 $ & $ -3.286 $ & $ -3.241 $ & $ 1.298 $ & $ -0.211 $ & $ 0.135 $ & $ -5.884 $ \\ \hline
\end{tabular}
  \label{tab:inferred_params_340} 
\end{center}
\end{table*}

\begin{table*}[]
\caption{Model parameters for distances between 464.2 and 1000.0 kpc.}
\begin{center}
\renewcommand{\arraystretch}{1.1}
\setlength{\tabcolsep}{11pt}
\begin{tabular}{c | c c c c c c c c}
\hline \hline \rule{0pt}{2.2ex}
	Configuration &  $a$ & $b$ & $\alpha$ & $M_{V, \mathrm{shift}}$ & $r_{h, \mathrm{shift}}$ & $k_1$ & $k_2$ & $k_3$ \\ \hline
	$(g,r)$ & $ -1.195 $ & $ -5.81 $ & $ -2.78 $ & $ -4.417 $ & $ 1.599 $ & $ -0.171 $ & $ 0.07 $ & $ -7.519 $ \\
	
	$(r,i)$ & $ -1.274 $ & $ -5.899 $ & $ -3.11 $ & $ -4.627 $ & $ 1.599 $ & $ -0.167 $ & $ 0.047 $ & $ -7.686 $ \\
	
	$(g,r,i)$ & $ -4.681 $ & $ -0.942 $ & $ 1.839 $ & $ -4.343 $ & $ 1.694 $ & $ -0.181 $ & $ 0.067 $ & $ -7.299 $ \\
	
	$(u,g)$ & $ -0.71 $ & $ -4.338 $ & $ -2.053 $ & $ -4.286 $ & $ 1.697 $ & $ -0.342 $ & $ 0.183 $ & $ -4.897 $ \\
	
	$(u,g,r)$ & $ -0.585 $ & $ -4.224 $ & $ -2.469 $ & $ -4.685 $ & $ 1.698 $ & $ -0.345 $ & $ 0.257 $ & $ -4.595 $ \\ 
	
	$(u,g,r,i)$ & $ -4.765 $ & $ -0.503 $ & $ 3.223 $ & $ -4.771 $ & $ 1.599 $ & $ -0.344 $ & $ 0.313 $ & $ -4.435 $ \\ \hline
\end{tabular}
  \label{tab:inferred_params_732} 
\end{center}
\end{table*}

\end{appendix}

\end{document}